\documentstyle[12pt]{article}
\def\bea{\begin{eqnarray}}
\def\eea{\end{eqnarray}}
\def\nn{\nonumber}
\def\beq{\begin{equation}}
\def\eeq{\end{equation}}
\def\ba{\beq\new\begin{array}{c}}
\def\ea{\end{array}\eeq}
\def\be{\ba}
\def\ee{\ea}
\def\stackreb#1#2{\mathrel{\mathop{#2}\limits_{#1}}}

\makeatletter
\newdimen\normalarrayskip              
\newdimen\minarrayskip                 
\normalarrayskip\baselineskip
\minarrayskip\jot
\newif\ifold             \oldtrue            \def\new{\oldfalse}
\def\arraymode{\ifold\relax\else\displaystyle\fi} 
\def\eqnumphantom{\phantom{(\theequation)}}     
\def\@arrayskip{\ifold\baselineskip\z@\lineskip\z@
     \else
     \baselineskip\minarrayskip\lineskip2\minarrayskip\fi}
\def\@arrayclassz{\ifcase \@lastchclass \@acolampacol \or
\@ampacol \or \or \or \@addamp \or
   \@acolampacol \or \@firstampfalse \@acol \fi
\edef\@preamble{\@preamble
  \ifcase \@chnum
     \hfil$\relax\arraymode\@sharp$\hfil
     \or $\relax\arraymode\@sharp$\hfil
     \or \hfil$\relax\arraymode\@sharp$\fi}}
\def\@array[#1]#2{\setbox\@arstrutbox=\hbox{\vrule
     height\arraystretch \ht\strutbox
     depth\arraystretch \dp\strutbox
     width\z@}\@mkpream{#2}\edef\@preamble{\halign
\noexpand\@halignto
\bgroup \tabskip\z@ \@arstrut \@preamble \tabskip\z@ \cr}%
\let\@startpbox\@@startpbox \let\@endpbox\@@endpbox
  \if #1t\vtop \else \if#1b\vbox \else \vcenter \fi\fi
  \bgroup \let\par\relax
  \let\@sharp##\let\protect\relax
  \@arrayskip\@preamble}
\def\eqnarray{\stepcounter{equation}%
              \let\@currentlabel=\theequation
              \global\@eqnswtrue
              \global\@eqcnt\z@
              \tabskip\@centering
              \let\\=\@eqncr
              $$%
 \halign to \displaywidth\bgroup
    \eqnumphantom\@eqnsel\hskip\@centering
    $\displaystyle \tabskip\z@ {##}$%
    \global\@eqcnt\@ne \hskip 2\arraycolsep
         $\displaystyle\arraymode{##}$\hfil
    \global\@eqcnt\tw@ \hskip 2\arraycolsep
         $\displaystyle\tabskip\z@{##}$\hfil
         \tabskip\@centering
    &{##}\tabskip\z@\cr}
\begingroup\ifx\undefined\newsymbol \else\def\input#1 {\endgroup}\fi
\newfont{\hr}{msbm10}
\newfont{\ams}{msam10}

\font\numbers=cmss12
\font\upright=cmu10 scaled\magstep1
\def\stroke{\vrule height8pt width0.4pt depth-0.1pt}
\def\topfleck{\vrule height8pt width0.5pt depth-5.9pt}
\def\botfleck{\vrule height2pt width0.5pt depth0.1pt}
\def\Zmath{\vcenter{\hbox{\numbers\rlap{\rlap{Z}\kern 0.8pt\topfleck}\kern
2.2pt
                   \rlap Z\kern 6pt\botfleck\kern 1pt}}}
\def\Qmath{\vcenter{\hbox{\upright\rlap{\rlap{Q}\kern
                   3.8pt\stroke}\phantom{Q}}}}
\def\Nmath{\vcenter{\hbox{\upright\rlap{I}\kern 1.7pt N}}}
\def\Cmath{\vcenter{\hbox{\upright\rlap{\rlap{C}\kern
                   3.8pt\stroke}\phantom{C}}}}
\def\Rmath{\vcenter{\hbox{\upright\rlap{I}\kern 1.7pt R}}}
\def\Z{\ifmmode\Zmath\else$\Zmath$\fi}
\def\Q{\ifmmode\Qmath\else$\Qmath$\fi}
\def\N{\ifmmode\Nmath\else$\Nmath$\fi}
\def\C{\ifmmode\Cmath\else$\Cmath$\fi}
\def\R{\ifmmode\Rmath\else$\Rmath$\fi}

\newcounter{app}

\def\app{\setcounter{equation}{0}
\def\theequation{\Alph{app}.\arabic{equation}}\par
   \addvspace{4ex}
   \@afterindentfalse
  \secdef\@app\@dapp}

\newcommand\@app{\@startsection {app}{1}{0ex}%
                                   {-3.5ex \@plus -1ex \@minus -.2ex}%
                                   {2.3ex \@plus.2ex}%
                                   {\normalfont\Large\bf}}
\def\@dapp#1{%
{\parindent \z@ \raggedright  \bf #1}\par\nobreak}
\def\l@app#1#2{\ifnum \c@tocdepth >\z@
    \addpenalty\@secpenalty
    \addvspace{1.0em \@plus\p@}%
    \setlength\@tempdima{8em}%
    \begingroup
      \parindent \z@ \rightskip \@pnumwidth
      \parfillskip -\@pnumwidth
      \leavevmode \bfseries
      \advance\leftskip\@tempdima
      \hskip -\leftskip
      #1\nobreak\hfil \nobreak\hb@xt@\@pnumwidth{\hss #2}\par
    \endgroup\fi}
\newcounter{sapp}[app]

\def\sapp{\def\theequation{\Alph{app}.\arabic{equation}}
\par
\@afterindentfalse
  \secdef\@sapp\@dsapp}
\newcommand{\@sapp}{\@startsection{sapp}{2}{\z@}%
                                     {-3.25ex\@plus -1ex \@minus -.2ex}%
                                     {1.5ex \@plus .2ex}%
                                     {\normalfont\large\bfseries}}

\def\@dsapp#1{%
{\parindent \z@ \raggedright  \bf #1
}\par\nobreak}
\newcommand{\l@sapp}{\@dottedtocline{2}{1.5em}{2.3em}}

\def\stackreb#1#2{\mathrel{\mathop{#2}\limits_{#1}}}

\def\2{{1\over 2}}
\def\N2{${\cal N}=2$}

\def\be{ \begin{eqnarray} }
\def\ee{ \end{eqnarray} }

\def\bea{\begin{eqnarray}}
\def\eea{\end{eqnarray}}
\def\nn{\nonumber}

\def\beq{\begin{equation}}
\def\eeq{\end{equation}}
\def\ba{\beq\new\begin{array}{c}}
\def\ea{\end{array}\eeq}
\def\be{\ba}
\def\ee{\ea}
\def\stackreb#1#2{\mathrel{\mathop{#2}\limits_{#1}}}

\def\R{Ruijsenaars-Schneider\ }
\def\SW{Seiberg-\-Witten theory\ }
\def\W{Weierstrass\ }

\begin{document}
\begin{flushright}
\hfill ITEP/TH-70/98\\
\hfill FIAN/TD-06/98\\
hepth/9902205
\end{flushright}
\vspace{0.5cm}
\begin{center}
{\LARGE \bf The Ruijsenaars-Schneider Model in the Context of
Seiberg-Witten
Theory}
\vspace{0.5cm}

\setcounter{footnote}{1}
\def\thefootnote{\fnsymbol{footnote}}
{\Large H.W.Braden\footnote{Department of Mathematics and Statistics,
University of Edinburgh, Edinburgh EH9 3JZ Scotland;
e-mail address: hwb@ed.ac.uk },
A.Marshakov\footnote{Theory
Department, Lebedev Physics Institute, Moscow
~117924, Russia; e-mail address: mars@lpi.ac.ru}\footnote{ITEP,
Moscow 117259, Russia; e-mail address:
andrei@heron.itep.ru},
A.Mironov\footnote{Theory
Department, Lebedev Physics Institute, Moscow
~117924, Russia; e-mail address: mironov@lpi.ac.ru}\footnote{ITEP,
Moscow 117259, Russia; e-mail address:
mironov@itep.ru}, A.Morozov\footnote{ITEP, Moscow
117259, Russia; e-mail address: morozov@vx.itep.ru}
}\\
\end{center}
\bigskip
\begin{quotation}
The compactification of five dimensional $N=2$ SUSY Yang-Mills (YM)
theory onto a circle provides a four dimensional YM model with
$N=4$ SUSY. This supersymmetry can be broken down to $N=2$ if
 non-trivial boundary conditions in the compact dimension,
$\phi(x_5 +R) = e^{2\pi i\epsilon}\phi(x_5)$, are imposed on half of
the fields.
This two-parameter $(R,\epsilon)$ family of compactifications
includes as particular limits most of the previously studied
four dimensional $N=2$ SUSY YM models with supermultiplets
in the adjoint representation of the gauge group.
The finite-dimensional integrable system associated to these theories
via the Seiberg-Witten construction is the generic elliptic
\R model. In particular the perturbative (weak coupling) limit is
described by the trigonometric \R model.
\end{quotation}

\section{Introduction}
\setcounter{footnote}{0}
\setcounter{equation}{0}

\SW \cite{SW} provides a description of the effective
low-energy actions of four dimensional $N=2$ SUSY Yang-Mills theories
in terms of finite-dimensional integrable systems \cite{GKMMM,rev}.
Such a description has been extended to five dimensional $N=2$ SUSY
theories \cite{N,S} with  one dimension compactified on
a circle of radius $R$ \cite{N}.
By starting with such a five dimensional model
one may obtain four dimensional $N=2$ SUSY models (with fields only
in the adjoint representation of the gauge group)
by imposing non-trivial boundary conditions on half of the fields:
\be
\phi(x_5 +R) = e^{2i\epsilon}\phi(x_5).
\label{bc}
\ee
If $\epsilon = 0$ one obtains $N=4$ SUSY in four dimensions,
but when $\epsilon \neq 2\pi n$ this is explicitly broken to $N=2$.
The low-energy mass spectrum of the  four dimensional theory contains
two towers of Kaluza-Klein modes:

\be
M = \frac {\pi n}{R}\ \ {\rm and} \ \
M = \frac{\epsilon + \pi n}{R}, \ \ \ N\in { Z}.
\label{spectrum}
\ee
In accordance with \cite{N,WDVV2} the prepotential ${\cal F}(a_i)$
for the group $SU(N)$ ($i = 1\ldots N$) should be
\be\label{pp}
\begin{array}{rl}
{\cal F}(a)&=-\tau\sum_i^Na_i^2+{1\over i\pi}
\left(\frac{ 1}{R }\right)^2\
\sum_{i,j}\hspace{-0.2cm}{\phantom{a}}^{'} \sum_{n=-\infty}^\infty
\Bigg\lbrace \left(Ra_{ij}+\pi n \right)^2 \log\left(Ra_{ij}+\pi n\right)\\
&\qquad -\left(Ra_{ij}+\pi n-\epsilon\right)^2
\log\left(Ra_{ij}+\pi n-\epsilon\right)+\hbox{regulator}\Bigg\rbrace\\
&\qquad +\hbox{non-perturbative (instanton-induced)  corrections}\\
&=\sum_{i,j}\hspace{-0.2cm}{\phantom{a}}^{'}f(a_{ij}) + \hbox{corrections}.
\end{array}
\ee
Here
\be
f(a)=-\frac{\tau}{N}a^2+\frac{1}{2i\pi R^2}\left(
\hbox{Li}_3\left(e^{-2iRa}\right)
-\hbox{Li}_3\left(e^{-2i(Ra+\epsilon)}\right)\right)
\ee
and
\be
\begin{array}{rl}
\frac{\partial^2 f}{\partial a^2}
&=-\frac{2\tau}{N} + {1\over i\pi}\sum_{n = -\infty}^{\infty} \left(
\log \left(R a + {\pi n}\right)
- \log \left(R a + {\epsilon + \pi n}\right)
\right) + \hbox{corrections} \\& =
\frac{-2\tau}{N} +
{1\over i\pi}\log\frac{\sin aR}{\sin (aR + \epsilon)} +
\hbox{corrections}.
\end{array}
\label{prep}
\ee
In these formulas $\tau = \frac{4\pi i}{g^2} + \frac{\theta}{2\pi}$
is the bare coupling constant and
$a_{ij} \equiv a_i - a_j$, where $ a_i$ ($\sum_i^N a_i=0$) characterize the
(eigenvalues of the) vacuum expectation values of the scalar fields
which spontaneously break the $SU(N)$ gauge symmetry
down to the  abelian one,  $U(1)^{N-1}$. Note that $T_{ij}(a)\equiv
\frac{\partial^2 {\cal F}}{\partial a_i\partial a_j}$ are couplings
of the effective low-energy abelian theory and they depend on the choice of
vacuum. Such an explicit occurence of the bare coupling $\tau$ is typical of
UV-finite YM theories which possess the highest possible
supersymmetry, perhaps softly broken.

Now \SW  (implicitly) provides exact expressions for the
full non-perturbative prepotential ${\cal F}(a)$ (i.e. with all the
instantonic corrections included) via\footnote{
Our conventions are to include a factor of $1/(2\pi i)$ in the
definition of $\oint$. Thus $\oint dz/z =1$.}
\be
a_i = \oint_{A_i} dS, \nn \\
\frac{\partial{\cal F}}{\partial  a_i} = \oint_{B_i} dS.
\label{oints}
\ee
Here $dS$ is the generating 1-form on the spectral curve
${\cal C}$ of the associated one dimensional integrable model, and
$A_i$, $B_i$ are the conjugate 1-cycles on ${\cal C}$,
$A_i \circ B_j = \delta_{ij}$.
The parameters $a_i$ may be considered as (some) moduli of
the complex structures on ${\cal C}$.

According to the proposal of \cite{N}, the prepotential (\ref{prep})
arising from this five dimensional theory may be
associated  with the elliptic \R integrable model \cite{R}.
In various double-scaling limits it reduces to well-studied systems:

(a) If $R\rightarrow 0$ (with finite $\epsilon$) the
(finite) mass spectrum (\ref{spectrum}) reduces to a single point
$M = 0$. This is the standard four dimensional $N=2$ SUSY YM model
-- the original and simplest example of \SW \cite{SW} --
associated with the periodic $A_{N-1}$ Toda chain
for $SU(N)$ \cite{GKMMM}
and with the appropriate periodic Toda chain for other gauge groups \cite{MW1}.
In this situation $N=2$ SUSY in four dimensions is insufficient to ensure
UV-finiteness, thus $\tau \rightarrow i\infty$, but the phenomenon of
dimensional transmutation occurs whereupon one
substitutes the dimensionless $\tau$ by
the new dimensionful parameter $\Lambda^N =
e^{2\pi i\tau}  (\epsilon/R)^{N}$.

(b) If $R \rightarrow 0$ and $\epsilon \sim mR$ for finite $m$,
then UV finiteness is preserved. The mass spectrum (\ref{spectrum})
reduces to the two points $M = 0$ and $M = m$. This is the
four dimensional YM model with $N=4$ SUSY softly broken to $N=2$.
The associated finite-dimensional integrable system \cite{DW,IM}
is the elliptic Calogero model \cite{Cal,KriF}.
Case (a) is then obtained from (b) by
Inosemtsev's \cite{Ino} double scaling limit
when $m\rightarrow 0$, $\tau \rightarrow i\infty$ and
$\Lambda^N = m^N e^{2\pi i\tau}$ is fixed.

(c) If $R \neq 0$ but $\epsilon \rightarrow i\infty$ the
mass spectrum (\ref{spectrum}) reduces to a single
Kaluza-Klein tower, $M = \pi n/R$, $n\in Z$. This compactification of
the five dimensional model has $N=1$ SUSY and is not
UV-finite. Here $\tau \rightarrow i\infty$ and
$\epsilon \rightarrow i\infty$, such that $2\pi\tau - N\epsilon$
remains finite. The corresponding integrable system \cite{N} is the relativistic
Toda chain \cite{relToda}.

(d) Finally, when $R\neq 0$ and $\epsilon$ and $\tau$
are both finite one distinguished case still remains:
$\epsilon =\pi/2$.\footnote{The case $\epsilon = 0$
of fully unbroken five dimensional $N=2$ supersymmetry
is of course also distinguished, but trivial:
there is no evolution of effective couplings (renormalisation
group flows) and the integrable system is just that of $N$
non-interacting (free) particles.
}
Here only periodic and antiperiodic boundary conditions
occur in the compact dimension. This is the
case analysed in \cite{BMMM1}. It is clearly special
from both the point of view of Yang-Mills theory and integrable
systems.

The purpose of the present paper is to provide further
details about the general \R prepotential (\ref{prep})
with arbitrary $R$ and $\epsilon$, as well as to show
what is special about the point $\epsilon =\pi/2$.
We shall mostly concentrate on the perturbative (weak
coupling) limit $\tau \rightarrow i\infty$ (with $R$
and $\epsilon$ fixed) when the instantonic corrections in
(\ref{prep}) can be neglected.
In contrast with the well-studied cases (a) and (c) the
perturbative limit for Calogero and \R systems at first sight appears
non-trivial: here elliptic systems are reduced to their
trigonometric counterparts and the spectral curve appears
complicated. As we shall show, however, the spectral curve may be
recast in rational form.\footnote{
In \cite{WDVV2, DP} the perturbative period matrix
$$
\frac{\partial^2{\cal F}}{\partial a_i\partial a_j} =
\log \frac{a_{ij}}{a_{ij}+m}
$$
was deduced from the elliptic case by ingenious yet
tedious calculations, which did not make explicit use
of the fact that elliptic Calogero model can be
substituted by its trigonometric limit (the Sutherland
model).}
In contrast to earlier works, this characterisation of the curve
enables us to give the first calculation of the perturbative
prepotential.
In the perturbative limit of case (d) we reobtain
the curve considered in \cite{BMMM1}.

An outline of the paper is as follows. First we will briefly review
the basics of Seiberg-Witten theory and the \R model. These sections
provide enough information to make the paper self-contained, and we
reference works that examine these topics more extensively.
Our strategy is to calculate the third derivatives of the
prepotential via information about the spectral curves of the
models. These curves are complicated, and section 4 of the paper
treats the case of $SU(2)$ as an illustrative example.
After this we treat the case of general $SU(N)$. One of our important
results is that the spectral curves admit a nice separation of
variables in the perturbative limit, which allows for
calculation.

\section{Basics of SW theory \label{basSW}}

A key feature relating SW theory and the theory of integrable
systems --and one still lacking a complete explanation-- is the
following:
the families of spectral curves $\{{\cal C}\}$ arising from SW
theory are parameterized by the Hamiltonians of an
associated integrable system.
For a given group and model the genus of ${\cal C}$ is fixed
and our family of curves is a certain subspace (normally of high
codimension) in the moduli space of complex structures.
An explicit parameterisation of curves is provided with the help of Lax
operator of integrable system. For the group $SU(N)$ this has
the form of an $N\times N$  matrix\footnote{
There may also be other representations. For example,
in terms of the $2\times 2$ transfer matrix
$T_N(\lambda)$ of some spin-chain model:
$$
{\det}_{2\times 2}  \left(f(\xi) I - T_N(\lambda)\right) = 0.
$$
Such alternative representations are crucially  important for the
description of matter in the fundamental representation of the gauge group
\cite{fund}.
The exact (duality) relation between these $N\times N$ and $2\times 2$
representations still remains obscure and is beyond the
scope of the present paper.
}
${\cal L}$ given on \lq\lq the bare spectral curve" ${\cal E}$, and

\be
{\cal C}:\ \ \ {\det}_{N\times N}\left(\lambda I - {\cal L}(\xi)
\right) = 0.
\label{sc}
\ee
The curve is an $N$-fold covering over ${\cal E}$
parameterized by the spectral parameter $\xi$.

For YM theories which are UV-finite
the bare curve is elliptic (a torus) with its own modulus $\tau$ and
we choose a \lq\lq bare co-ordinate" $\xi$ so that
$d\xi$ is a holomorphic differential on ${\cal E}$.
In the double-scaling limits which lead to the UV-infinite theories
described above this modulus disappears and is  replaced by
the dimensionful $\Lambda$. In this situation the
bare spectral curve degenerates
into the doubly punctured Riemann sphere with spectral parameter
\be
w \sim e^{2 i\xi},
\label{wdef}
\ee
and $d\xi = \frac{1}{2 i}\frac{dw}{w}$.
The Lax operators and integrable systems we encounter reflect the
dimensionality of the underlying YM theories.
For five dimensional YM theories the Lax operators are \lq\lq
group-like"  objects ( with integrable systems the
\R and relativistic Toda models) while
for four dimensional theories they are \lq\lq algebra-like"  objects
(with integrable systems the Calogero and ordinary Toda chain).
The generating 1-forms are then found to be
\be
\begin{array}{rll}
dS &= R^{-1}\log\lambda d\xi\ & {\rm for}\ 5d\ {\rm models},\\
dS &= \lambda d\xi  &  {\rm for}\ 4d\ {\rm models}.
\end{array}
\label{dS}
\ee

Suppose we parameterize a bare curve that is elliptic by the
algebraic equation,
\be
\hat y^2
=\hat x^3 - \alpha \hat x^2 - \beta \hat x - \gamma.
\label{elc}
\ee
Then with $\hat x = \wp(\xi)+\alpha/3$ and
$\hat y= -\frac{1}{2}\wp'(\xi)$ we have that
$d\xi =-2\frac{d\hat x}{\hat y}$.  Here $\wp(\xi)$ is the
doubly periodic Weierstrass $\wp$-function with
periods $2\omega$ and $2\omega'$, and $\tau={\omega'\over\omega}$
\cite{BE, WW}. This satisfies the canonical equation
$$\wp'(\xi)\sp2 =4(\wp(\xi)- e_1) (\wp(\xi)-e_2)
(\wp(\xi)- e_3),\qquad e_1+e_2+e_3=0.$$
We remark that although the Weierstrass function depends on two
periods, the homogeneity relation
$$ \wp(t z|t\omega,t \omega')= t\sp{-2} \wp(z|\omega,\omega')$$
enables us to arbitrarily rescale one of these.
Our final results are independent of such scaling and this allows us to
choose the real period to be $\pi$ (that is $\omega=\pi/2$). The
perturbative limit is then given by the imaginary period
$2\omega'$ becoming infinite, $\tau \rightarrow i\infty$.
In this weak coupling limit (without any double-scaling) the
bare curve (\ref{elc}) becomes
\be
y^2 = x^2(x-1).
\label{perelc}
\ee
Then $\hat x\rightarrow x-1/3$ with $x = \frac{1}{\sin^2\xi}$,
$\hat y\rightarrow y = -\frac{\cos\xi}{\sin^3\xi}$,
$\alpha = 1$, $\beta = \gamma = 0$ and
$d\xi\rightarrow {dx\over{x\sqrt{x-1}}}$.
Throughout we will denote  by \lq\lq hatted" quantities those
expressions depending on an elliptic bare curve (such as $\hat x $)
while \lq\lq unhatted" quantities will denote their degenerations
in the $\tau \rightarrow i\infty$ limit (such as $x$).

The variations of $dS$ with respect to moduli are, by
definition, {\it holomorphic}
1-forms on the spectral curve. Particular choices of coordinates
for the moduli may have natural properties. For example,
if $a_i$ defined by (\ref{oints}) are chosen as
coordinates on the moduli space, then
\be
\frac{\partial dS}{\partial a_i} = d\Omega_i
\label{derdS}
\ee
are {\it canonical} 1-differentials,
$\oint_{A_i} d\Omega_j=\delta_{ij}$. When the prepotential
satisfies the WDVV equations this choice is equivalent to
specifying a flat structure \cite{dub, WDVV}. Further the second
derivatives $T_{ij}$ of the prepotential (\ref{prep}) with
respect to these coordinates then form
the period matrix of ${\cal C}$. An even simpler
expression -- the \lq\lq residue formula" -- exists for the variation
of the period matrix, that is for the third derivatives
of ${\cal F}(a)$ (see, e.g. \cite{WDVV2}):
\be
\frac{\partial^3{\cal F}}{\partial a_i\partial a_j\partial a_k}
= \frac{\partial T_{ij}}{\partial a_k} ={1\over 2\pi i}
{\rm res}_{d\xi = 0} \frac{d\Omega_id\Omega_jd\Omega_k}{\delta dS},
\label{res}
\ee
where $\delta dS\equiv d\left({dS\over d\xi}\right)d\xi $, or, explicitly
\be
\begin{array}{rll}
\delta dS &= R^{-1}d\log\lambda d\xi &{\rm for}\ 5d\ {\rm models},\\
\delta dS &= d\lambda d\xi &  {\rm for}\ 4d\ {\rm models}.
\end{array}
\label{ddS}
\ee
We remark that although
$d\xi$ does not have zeroes on the {\it bare} spectral
curve when it is a torus or doubly punctured sphere, it does
in general
however possess them on the covering ${\cal C}$. Examples of this
will be given later in the paper.

\section{Basics of the \R model}

The \R model is a remarkable completely integrable system whose
various limits include the (finite dimensional)
Toda and Calogero-Moser models.
Here we shall review only a few of its salient features needed
for our calculations. More comprehensive accounts of its structure
and applications may be found in \cite{ruijrev}.
The $GL(N)$ model has an explicit Lax representation with
Lax operator \cite{R, BC, RL}
\be
{\cal L}_{ij} = c(\xi|\epsilon)
e^{P_i} \frac{F(q_{ij}|\xi)}{F_(q_{ij}|\epsilon)}.
\label{RLa}
\ee
Here
\be
e^{P_i} = e^{p_i}
\prod_{l\neq i} \sqrt{\wp (q_{il})-\wp(\epsilon)}.
\label{RLb}
\ee
where the $p_i$ and $q_i$ are canonically conjugate momenta and
coordinates, $\{p_i, q_j\} = \delta_{ij}$.
The commuting Hamiltonians may be variously written
\be
 H_k =
\sum_{{J\subset \{1,\ldots n\}}\atop{ |J|=k}}
e\sp{\sum_{j\in J} p_j}
\prod_{{j\in J}\atop{ k\in \{1,\ldots n\}\backslash J}}
\frac{F(q_{ij}|\xi)}{F_(q_{ij}|\epsilon)}=
\sum_{1\leq i_1<\ldots<i_k\leq N} e^{P_{i_1}+\ldots+P_{i_k}}
\prod_{a<b}\frac{1}{\wp(q_{i_ai_b})-\wp(\epsilon)}.
\label{Hams}
\ee
(The final product is taken to be unity in the case $k=1$.)
These Hamiltonians arise in the description of the spectral curve
(\ref{sc}).
The special functions that appear above are defined to be
\cite{KriF}
\be
F(q|\xi) = \frac{\sigma(q - \xi)}{\sigma(q)\sigma(\xi)},
\label{Ffun}
\ee
where $\sigma(\xi)$
denotes the Weierstrass $\sigma$-function \cite{BE, WW}. One identity
used throughout is
\be
\frac{\sigma(u-v)\sigma(u+v)}{\sigma\sp2(u)\sigma\sp2(v)}=
\wp(v)-\wp(u).
\label{wpaddn}
\ee
The integrability of the model depends on the functional
equations satisfied by $F$ \cite{R}; the connection with functional
equations and integrable systems is part of a larger story \cite{funl}.

Some comment on the parameters $\xi$ and $\epsilon$ in these
formulae is called for. Here $\xi$ is precisely the spectral parameter
we have encountered in (\ref{sc}). Further, the
additional parameter $\epsilon$ in (\ref{RLa},\ref{RLb}) above
is the  same parameter we introduced in (\ref{bc})
characterising the boundary conditions.
Actually the integrability of  (\ref{Hams}) does not require
$\epsilon$ to be real, but such a choice guarantees the reality of the
the Hamiltonians. The identification of these two parameters has been
simplified by our choice of the real period of $\wp$ being $\pi$
so that both (\ref{bc}) and (\ref{RLa},\ref{RLb}) are
manifestly $\pi$ periodic.
The \lq\lq non-relativistic limit" $\epsilon \rightarrow 0$
which leads to the Calogero-Moser system means we can identify the
mass $m=\epsilon/R$ of the gauge multiplet with the Calogero-Moser
coupling constant.
The special point $\epsilon = \pi/2 $ ($=\omega$)
singled out earlier is now a half-period of the $\wp$-function,
and at this point $\wp'(\frac{\pi}{2}) = 0$.

>From the point of view of classical integrability the
overall normalisation factor $c(\xi|\epsilon)$ of the Lax operator
does not change the integrals of motion apart from scaling.
This normalisation factor does however lead to the rescaling of
$\lambda$ and this effects the explicit form of
the 1-form $dS$.\footnote{
Note that although the choice of  normalisation $c(\xi|\epsilon)$
(or, equivalently, rescaling $\lambda$)
can change the manifest expression
for $dS$, it affects neither the symplectic 2-form
${d\lambda\over\lambda}\wedge{dw\over w}$,
nor the period matrix, which is the second derivative of the
prepotential.
}
 We shall see below
that significant simplifications occur with the choice
\be
c(\xi|\epsilon) =
c_0(\xi|\epsilon) \equiv
 \frac{1}{\sqrt{ \wp(\xi)-\wp (\epsilon)}}=
\frac{\sigma(\xi)\sigma(\epsilon)}{\sqrt{\sigma(\epsilon-\xi)
\sigma(\epsilon+\xi)}} = \sqrt{c_+(\xi|\epsilon)c_-(\xi|\epsilon)}
\label{cfun}
\ee
where
\be
c_{\pm}(\xi|\epsilon)\equiv \frac{\sigma(\xi)\sigma(\epsilon)}
{\sigma(\epsilon\pm\xi)}=\mp\frac{1}{F(\epsilon|\mp\xi)}.
\ee
Similar issues of normalisation enter into discussion of
separation of variables \cite{separation}.

For future reference it is convenient to record here
the value of these expressions in
the perturbative (weak coupling) limit $\tau \rightarrow i\infty$.
We have already seen  that
$ \wp(\xi) \rightarrow \frac{1}{\sin^2 \xi}-{1\over 3} $.
Further,
\be\label{ctrig}
c_0(\xi|\epsilon)\rightarrow{1\over\sqrt{\sin^{-2}\xi-
\sin^{-2}\epsilon}}=
\frac{\sin\xi\sin\epsilon}{\sqrt{\sin(\epsilon-\xi)
\sin(\epsilon+\xi)}},\ \ \
c_{\pm}(\xi|\epsilon) \rightarrow
\frac{\sin\xi\sin\epsilon}
{\sin(\epsilon\pm\xi)}e^{\mp\epsilon\xi/3}\equiv
\bar c_\pm e^{\mp\epsilon\xi/3}.
\ee
The corresponding integrable system is  then nothing but the
trigonometric \R model.
Finally the special point $\epsilon = \omega = \frac{\pi}{2}$
sees
\be
\wp(\omega) = e_1 \rightarrow  {2\over 3}
\ee
and the particular combination (\ref{cfun}) becomes
\be
c_0(\xi|\epsilon =\pi/2) \rightarrow \tan \xi.
\label{cfunp}
\ee

\section{The case of SU(2)}

Before generalising to the higher rank situation it is instructive
to first consider the case of gauge group SU(2).
In the formulae of the previous section this corresponds to
setting $N=2$ and working in the center-of-mass frame $p_1+p_2=0$.
We define $p\equiv p_1$ and $q\equiv q_1-q_2$.
Having constructed the spectral curve we shall proceed to calculate
the third derivative of ${\cal F}(a)$, first directly, then via
the residue formula. These will be seen to agree with those
coming from (\ref{prep}).

\subsection{Spectral curve}

Using the explicit form of the Lax operator the
spectral curve (\ref{sc}) is found to be
\be
\lambda^2 - cu\lambda + c^2(\wp(\xi) - \wp(\epsilon)) =
\lambda^2 - cu\lambda + \frac{c^2}{c_0^2} = 0,
\label{SU2c}
\ee
where
\be
u\equiv  H_1(p,q)=2\cosh p\sqrt{\wp (q)-\wp (\epsilon)}
\ee
is the  Hamiltonian (\ref{Hams}).
For the choice   $c(\xi|\mu) = c_0(\xi|\mu)$ this simplifies to yield
\be
\lambda^2 - uc_0 \lambda + 1 = 0,
\ee
or simply
\be
c_0^{-1}(\xi) = \frac{u\lambda}{\lambda^2 + 1} =
\frac{u}{\lambda + \lambda^{-1}}.
\label{SU2c1}
\ee
Observe that with our choice (\ref{cfun}) this equation,
which describes the Seiberg-Witten spectral curve, may be expressed
in the form.
\be
u= H(\log\lambda,\xi).
\ee
Comparison with (\ref{dS}) shows that the generating differential
$R dS=\log\lambda d\xi$ takes the form $pdq$ here.
This appears to be the case in all known examples \cite{GKMMM}:
the Seiberg-Witten generating differential takes the form
$pdq$ for the corresponding integrable system. A general proof of this
rather natural correspondence is still lacking.

Upon using (\ref{cfunp}) we see that (\ref{SU2c1}) reduces to
\be
\cot\xi = \frac{u\lambda}
{\lambda^2 + 1}
\label{SU2cot}
\ee
at the special point $\epsilon =\pi/2$ and in the perturbative limit.
The rational spectral curve for this situation was given
in \cite{BMMM1}\footnote{This paper in fact used the slightly different
normalisation $w=-2e^{2i\xi}$ from that being used here and we have
compensated for this in the expressions being presented.} as
\be
w = \frac{(\lambda - \lambda_1)(\lambda - \lambda_1^{-1})}
{(\lambda + \lambda_1)(\lambda + \lambda_1^{-1})}.
\label{SU2B}
\ee
These two expressions are seen to agree upon setting
$w = -e^{2 i\xi}$ and $i u = (\lambda_1 + \lambda_1^{-1})$.

\subsection{The period matrix in the perturbative limit}

As we described in section \ref{basSW}, the derivatives of
the 1-form $dS$ with respect to the
moduli are holomorphic differentials.  A particular choice of
coordinates for the moduli will lead to the canonical  holomorphic
1-differentials. For the example at hand there is only one modulus and
\be
dv = \frac{\partial dS}{\partial u} =
R^{-1}\frac{1}{\lambda}\frac{\partial\lambda}{\partial u} d\xi
= R^{-1}\frac{c_0d\xi}{2\lambda - c_0u} =
{1\over R}\frac{d\xi}{\sqrt{u^2 - 4/c_0^2}}
={1\over{ 2 R}}\frac{d\xi}{\sqrt{\frac{u^2}{4} +
\wp(\epsilon)-\wp(\xi)}}.
\ee
If we order the roots  $e_3\le e_2\le e_1$  we may take
for the $A$ integral
\be
\oint_A dv = {1\over i\pi}\int_{e_3}\sp{ e_2} dv
=\frac{1}{i\pi R \sqrt{( \frac{u^2}{4} +\wp(\epsilon)-e_2)(e_1-e_3)}}
K\left(\sqrt{\frac{
(\frac{u^2}{4} +\wp(\epsilon)-e_1)(e_2-e_3)
                  }
{(\frac{u^2}{4} +\wp(\epsilon)-e_2)(e_1-e_3)
                  }
             }
 \,\right).
\label{intAdvtorus}
\ee
Here $K(q)$ is the complete elliptic integral of the first kind.
Dividing $dv$ by the right hand side of this expression would then
give us the canonical holomorphic 1-differential $d\Omega$.
Similarly for the $B$ integral we have
\be
\oint_B dv = {1\over i\pi}\int_{e_2}\sp{ e_1} dv
=-\frac{1}{\pi R \sqrt{( \frac{u^2}{4} +\wp(\epsilon)-e_2)(e_1-e_3)}}
K\left(\sqrt{\frac{
(\frac{u^2}{4} +\wp(\epsilon)-e_3)(e_1-e_2)
                  }
{(\frac{u^2}{4} +\wp(\epsilon)-e_2)(e_1-e_3)
                  }
             }
 \,\right),
\ee
and so the period matrix is
\be
T=\oint_B d\Omega=
-i K\left(\sqrt{\frac{
(\frac{u^2}{4} +\wp(\epsilon)-e_3)(e_1-e_2)
                  }
{(\frac{u^2}{4} +\wp(\epsilon)-e_2)(e_1-e_3)
                  }
             }
 \,\right)
\Bigg/
K\left(\sqrt{\frac{
(\frac{u^2}{4} +\wp(\epsilon)-e_1)(e_2-e_3)
                  }
{(\frac{u^2}{4} +\wp(\epsilon)-e_2)(e_1-e_3)
                  }
             }
 \,\right).
\label{intBdvtorus}
\ee
One can take the limit $\epsilon\to 0$ in this expression to
obtain $T=-\tau$. This agrees perfectly with formula (\ref{prep})
for $SU(2)$ and justifies our identification of the gauge theory
coupling constant and the modulus of the bare spectral curve.

Let us now calculate the same quantities in the perturbative limit.
Now $d\xi \rightarrow \frac{dx}{x\sqrt{x-1}}$,
$\wp(\xi)\rightarrow \sin^{-2}\xi -1/3$ and upon substituting
$1/c_0^2 = x - \sin^{-2}\epsilon$ from (\ref{cfun}) we obtain
\be
dv \rightarrow \left(\frac{1}{2R}\right)
\frac{dx}{x\sqrt{(x-1)(U^2 - x)}},
\ee
with $U^2 = \frac{1}{\sin^2\epsilon} + \frac{u^2}{4}$.
We have $e_2=e_3=-1/3$ and $e_1=2/3$ in this limit and so
the $A$-period in this case shrinks to a contour around $x=0$. Now
$\oint_A dv = 1/(2iRU)$ and we may identify the
canonical differential $d\Omega=2iRUdv$.
This result also follows from (\ref{intAdvtorus}) upon using
$$K(q)\stackreb{q\rightarrow 0}{=}
\frac{\pi}{2}(1+\frac{q^2}{4}+\ldots).
$$
The $B$-period of $d\Omega$ again
gives the period matrix. The corresponding
integral now goes (twice) between $x=0$ and
$x=1$, and the integral $\oint_B dv$  diverges logarithmically
in the vicinity of $x= 0$. This divergence was to be expected
because the period matrix (\ref{prep}) contains a term $\tau$ on the
right hand side  and the perturbative limit is given by
$\tau \rightarrow i\infty$.
Upon making the rational substitution $x=\frac{v}{1+v}$ we obtain
\be
\begin{array}{rl}
T&= \oint_B d\Omega=
{U\over\pi}
\int_0^1 \frac{dx}{x\sqrt{(x-1)(U^2-x)}} = \lim_{\varepsilon\to 0}
\frac{U}{i\pi\sqrt{U^2-1}}\int_\varepsilon^\infty
\frac{dv}{v\sqrt{v+ \frac{U^2}{U^2-1}}}
\\
&=-
{1\over i\pi}\lim_{\varepsilon\to 0}
\left(\log {\varepsilon\over 4}\right)+
{1\over i\pi}\log\frac{U^2}{1-U^2}
\end{array}
\label{int2}
\ee
where $\varepsilon$ is a small-$x$ cut-off. Thus,
the $U$ dependent part of this integral is finite and can be
considered as the \lq\lq true" perturbative correction, while
the divergent part just renormalises the bare  \lq\lq classical"
coupling constant $\tau$. Again the same result follows from
(\ref{intBdvtorus}) upon using
$$\lim_{q\rightarrow 1}
\bigg( K(q) -\frac{1}{2}\ln (\frac{16}{1-q\sp2})\bigg)=0.
$$

The final ingredient we wish are the $a$-variables,
i.e. the $A$-period of $dS$ itself. This will correspond to the
integral of (\ref{intAdvtorus}) with respect to $u$, which is a
rather complicated integral. For our purposes
the perturbative limit will suffice when there are several
simplifications. From the definition of (\ref{oints}) and that of
$dS$ we find
$$a=\frac{1}{ R}\oint\frac{\log \lambda dx}{x\sqrt{x-1}}
=\frac{\log\lambda|_{x=0}}{i R}.
$$
Now at $x=0$ we have that $c_0 =i \sin\epsilon$ while at
the same time $\lambda + \lambda^{-1} = c_0u$. Together these yield
$$2\cos aR=i u\sin\epsilon$$
and
\be
U^2 = \frac{u^2}{4} + \frac{1}{\sin^2\epsilon} =
\frac{1}{\sin^2\epsilon} (1 - \cos^2 aR) =
\frac{\sin^2aR}{\sin^2\epsilon}.
\ee
Substituting this expression for $U^2$ into (\ref{int2})
we obtain for the finite part of the period matrix
\be\label{su2pp}
T_{\rm{finite}}
 = {1\over i\pi}\log \frac{\sin^2 aR}{\sin^2\epsilon - \sin^2 aR}=
{1\over i\pi}\log\frac{\sin^2aR}{\sin\left(\epsilon+aR\right)
\sin\left(\epsilon -aR\right)}.
\ee
Upon using $\sin^2 a - \sin^2 b=\sin (a+b)\sin(a-b)$ this yields
precise agreement with (\ref{prep}) where in the case of $SU(2)$
we note that
only two terms contribute to the sum: one with $a_{12} = a$ and
one with $a_{21} = - a$.

\subsection{The perturbative residue formula}

We remark that the problems we encountered in the previous section of
divergent integrals  do not arise when working with the derivatives
of the period matrix with respect to its moduli (i.e. with the
third derivative of the prepotential).
Indeed the most effective way to deal with
perturbative prepotentials is to calculate them via residue formulae,
which are finite. Here we will illustrate the residue
formula (\ref{res}) for the system at hand, reproducing
the result (\ref{su2pp}) of the previous subsection.

We wish to calculate
\be
\frac{\partial^3{\cal F}}{\partial a^3}
= \frac{\partial T}{\partial a} ={1\over 2\pi i}
{\rm res}_{d\xi = 0} \frac{d\Omega d\Omega d\Omega}{\delta dS}.
\ee
>From the previous section we have the canonical differential
\be
d\Omega= iU\frac{dx}{x\sqrt{(x-1)(U^2 - x)}}.
\ee
Also from
$$ x=1/c_0^2+\sin^{-2}\epsilon={\lambda^2 u^2\over (\lambda^2 +1)^2}
+\sin^{-2} \epsilon=U\sp2 -\frac{u\sp2}{4}\left(\frac{1-\lambda\sp2}
{1+\lambda\sp2}\right)\sp2$$
we obtain that
$$
\frac{dx}{d\lambda}=
2\lambda u \frac{1-\lambda\sp2}{(1+\lambda\sp2)\sp3}
$$
and consequently that
\be
d\xi = \frac{dx}{x\sqrt{x-1}} ={d\lambda\over\sqrt{{\lambda^2
u^2\over (\lambda^2+1)^2}+\sin^{-2}\epsilon -1}}
{2\lambda u^2\over \lambda^2 u^2 +\sin^{-2}\epsilon (\lambda^2
+1)^2}{1-\lambda^2\over 1+\lambda^2}.
\label{dxdl}
\ee
Observe that although the expression for $d\xi$ is nonvanishing
as a function of the bare spectral parameter $x$ (for finite $x$),
it does however vanish as a function of the proper local parameter
$\lambda\sp2$ at $\lambda^2=1$ (when $x=U\sp2$). That is $d\xi$, while
nonvanishing as a function of the bare spectral curve,
does vanish on its cover ${\cal C}$.

Putting these various expressions together yields
\be
\frac{d\Omega d\Omega d\Omega}{\delta dS}=
\frac{i Ru U\sp3}{x\sp2 (x-1)}\frac{16\lambda\sp2}{(1+\lambda\sp2)\sp3}
\frac{d\lambda\sp2}{1-\lambda\sp2}.
\label{su2ress}
\ee
Now we can apply the residue formula (\ref{res}), which gives
\be
{\partial^3{\cal F}\over\partial a^3}=\frac{1}{\pi}{uR\over (1-U^2)U}.
\ee
This should be compared with the derivative of (\ref{su2pp})
with respect to $a$, that gives exactly the same result.

\section{The case of general $SU(N)$ \label{geng}}

We now consider the general $SU(N)$ model. The first step is to
evaluate the spectral curve (\ref{sc}) for the Lax matrix (\ref{RLa}).
Expanding the determinant about the diagonal yields
\be
\sum_{k=0}^N (-\lambda)^{N-k}c^k\left\{
\sum_{1\leq i_1<\ldots<i_k\leq N} e^{P_{i_1}+\ldots+P_{i_k}}
{\det}_{(ab)} \frac{F(q_{i_ai_b}|\xi)} {F(q_{i_ai_b}|\epsilon)}
\right\} = 0,
\label{Rsc1}
\ee
and Ruijsenaars \cite{R} expressed the determinants appearing here
in terms of the Hamiltonians (\ref{Hams}) by means of a generalised
Cauchy formula.
Here we shall reobtain this expansion using a simple fact that
generalised Cauchy formulae can be derived in terms of
free-fermion correlators via Wick's theorem for fermions.
The \R model (and its spin
generalisations) may be understood \cite{RL,R+T} in terms of a reduction
of the Toda lattice hierarchy. The Hirota bilinear identities
of that hierarchy may be expressed in terms of free-fermion
correlators \cite{DJKM},
and are the origin of those here. We believe that
these free field expansions will ultimately lead to a better
field theoretic understanding of the appearance of these integrable
systems. The machinery of free-fermion correlators has already found
use, for example, in calculating within the context of the
Whitham hierarchy for these integrable
systems \cite{RG}. Having obtained the spectral curve we will
then show that remarkable simplifications occur for the
parameterisation of the curve, both in the perturbative limit
and for the nonperturbative special point $\epsilon=\omega=\pi/2$.
It is the existence of these simpler rational forms
of the curves that enables us to calculate the prepotential.

\subsection{Handling determinants}

The determinants in (\ref{Rsc1}) can be evaluated by making use of
Wick's theorem for fermionic correlators on a Riemann surface.
(Appendix A of \cite{RG} provides both a summary and references
for this machinery.)
First observe that
\be
\frac{F(q_{ab}|\xi)} {F(q_{ab}|\epsilon)} =
\frac{\sigma(q_a - q_b - \xi)\sigma(\epsilon)}
{\sigma(q_a - q_b - \epsilon)\sigma(\xi)} \cong
\frac{\theta_e(q_a - q_b - \epsilon)}{\theta_*(q_a - q_b - \epsilon)
\theta_e(0)}\cdot\frac{\theta_*(\epsilon)\theta_*(\epsilon - \xi)}
{\theta_*(\xi)}.
\label{prel}
\ee
Here $*$  is the odd theta-characteristic (such that
$\theta_*(-z) = -\theta_*(z)$, traditionally labelled $\theta_1$ for
genus one), $e = \epsilon - \xi$ and
$\theta_e(z) \cong \theta_*(z+e)$. The symbol \lq\lq $\cong$"
denotes here equality modulo standard factors like quadratic
and linear exponents and Dedekind functions, which will
cancel in the final expressions. They may be simply restored upon using
$$\sigma(z)=\frac{2\omega}{\pi}e\sp{\frac{\eta z^2}{2\omega}}
\frac{\theta_1(\frac{\pi z}{2\omega})}{\theta_1\sp\prime(0)}.$$

Now the fermionic correlator on a torus is simply
$$
\Psi_e(\zeta|\eta)=\langle \psi(\zeta)\tilde\psi(\eta)\rangle_e
=\frac{1}{E(\zeta,\eta)}\frac{\theta_e(\zeta-\eta)}{\theta_e(0)}
=\frac{\theta_1\sp\prime(0)}{\theta_1(\zeta-\eta)}
\frac{\theta_e(\zeta-\eta)}{\theta_e(0)}\sqrt{d \zeta}\sqrt{d \eta},
$$
where $E(\zeta,\eta)$ is the prime form. With $\zeta=q_a$
and $\eta=q_b+ \epsilon$ we see that the first factor on the
right hand side of (\ref{prel}) is simply the fermionic correlator
$\Psi_e(q_a|q_b + \epsilon)$ on the torus. Further, the
multi-fermionic correlator also has a simple expression:
\be
\Psi_e(\zeta_1,\ldots,\zeta_k|\eta_1,\ldots, \eta_k) =
\ \langle \prod_{a=1}^k \psi(\zeta_a)\tilde\psi(\eta_a)\rangle_e
=
\frac{\prod_{a<b} E(\zeta_a,\zeta_b) E(\eta_a,\eta_b)}
{\prod_{a,b} E(\zeta_a,\eta_b)}\cdot
\frac{\theta_e(\sum_a \zeta_a - \sum_a \eta_a)}{\theta_e(0)}
\label{ferc}
\ee
Wick's theorem \cite{Wick} expresses the fact that such a
correlator has determinant form:
\be
{\det}_{(ab)} \Psi_e(\zeta_a|\eta_b) =
(-)^{k(k-1)\over 2}
\Psi_e(\zeta_1,\ldots,\zeta_k|\eta_1,\ldots, \eta_k).
\label{Wick}
\ee
(In fact (\ref{ferc}) (\ref{Wick}) hold
true for any genus, but here we only need them on the torus.)

We see then that (up to factors) the determinants
${\det}_{(ab)} \frac{F(q_{i_ai_b}|\xi)} {F(q_{i_ai_b}|\epsilon)}$
are the determinants of free fermion correlators, which may be
evaluated using (\ref{ferc}) and (\ref{Wick}).
Upon substituting $\zeta_a = q_{i_a}$, $\eta_a =
q_{i_a} + \epsilon$ and $e = \epsilon - \xi$,  and collecting
these factors we obtain the result of Ruijsenaars \cite{R}:
\be
\tilde D_{I_k} \equiv
{\det}_{(ab)} \frac{F(q_{i_ai_b}|\xi)} {F(q_{i_ai_b}|\epsilon)}
= (-)^{k(k-1)\over 2}
\frac{\sigma^{k-1}(\xi - \epsilon) \sigma(\xi + (k-1)\epsilon)}
{\sigma^k(\xi) \sigma^{k(k-1)}(\epsilon)}
\prod_{a<b} \frac{\sigma^2(q_{i_ai_b})\sigma^2(\epsilon)}
{\sigma(\epsilon+q_{i_ai_b}) \sigma(\epsilon-q_{i_ai_b})},
\label{tDk}
\ee
where $I_k = \{i_1,\ldots,i_k\}$ denotes the set of $k$ indices
and $a,b=1,...,k$.
The right hand side of this expression is in fact a doubly
periodic function of all the arguments ($q_i$, $\xi$ and $\epsilon$)
and so may rewritten in terms of \W functions.
In particular, upon using (\ref{wpaddn}), we find that
\be
\prod_{a<b} \frac{\sigma^2(q_{i_ai_b})\sigma^2(\epsilon)}
{\sigma(\epsilon +q_{i_ai_b}) \sigma(\epsilon -q_{i_ai_b})}
= \prod_{a<b} \frac{1}{\wp (q_{i_ai_b})-\wp (\epsilon))},
\ee
thus recovering the Hamiltonians (\ref{Hams}) noted earlier.

\subsection{Spectral curve \label{scsec}}

Upon substitution of (\ref{tDk}) into (\ref{Rsc1})
the \R spectral curve takes the form
\be
\sum_{k=0}^N (-\lambda)^{N-k}c^k\left\{
\sum_{I_k} e^{P_{i_1}+\ldots+P_{i_k}}\tilde D_{I_k}\right\} =
\sum_{k=0}^N (-\lambda)^{N-k} c^kD_k(\xi|\epsilon)\  H_k
=0.
\label{Rsc}
\ee
Here we have collected the $q$-independent factors in (\ref{tDk})
into $D_k$ where
\be
D_k(\xi|\epsilon) =
(-)^{k(k-1)\over 2}
\frac{\sigma^{k-1}(\xi - \epsilon) \sigma(\xi + (k-1)\epsilon)}
{\sigma^k(\xi) \sigma^{k(k-1)}(\epsilon)}
 =\frac{1}{c_-\sp{k}}. \frac{\sigma(\xi + (k-1)\epsilon)}{
\sigma(\xi - \epsilon)}.\frac{(-1)\sp{k(k-1)\over 2}}
{\sigma^{k(k-2)}(\epsilon)}.
\label{Dfun}
\ee
Again this is doubly periodic function in both $\xi$ and $\epsilon$
and so is expressible in terms of the \W function and
its derivative, though the explicit formulae are rather complicated.
Although it may appear from the last expression in
(\ref{Dfun}) that (\ref{Rsc}) simplifies when $c=c_-$, such a choice
would break the double periodicity of our spectral curve and so
is inappropriate for the fully elliptic model. However, in
the perturbative limit when one of the periods becomes infinite
this choice is then available.

Part of the difficulty in dealing with the elliptic \R
and Calogero-Moser models is the complicated nature of these
spectral curves.
For example, in the case of $SU(3)$ the spectral curve (\ref{sc}) is
\be
\lambda^3 - cu\lambda^2 + c^2v\lambda(\wp(\xi) - \wp(\epsilon)) +\\
+ c^3\left(\frac{1}{2}\wp'(\xi)\wp'(\epsilon)  - 3\wp(\xi)\wp^2(\epsilon)
+ \wp^3(\epsilon) + 2\alpha \wp(\xi)\wp(\epsilon) + 2\beta
(\wp(\xi) + \wp(\epsilon)) + 2\gamma\right)
= 0\\
u\equiv H_1=H_+,
v\equiv H_2=H_-
\label{SU3cc}
\ee
where
\be
H_{\pm}=e^{\pm p_1}\sqrt{\wp(q_{12})-\wp(\epsilon)}
\sqrt{\wp(q_{13})-\wp(\epsilon)} +
e^{\pm p_2}\sqrt{\wp(q_{12})-\wp(\epsilon)}
\sqrt{\wp(q_{23})-\wp(\epsilon)}+\\
+e^{\pm p_3}\sqrt{\wp(q_{13})-\wp(\epsilon)}
\sqrt{\wp(q_{23})-\wp(\epsilon)},\ \ \ \ p_1+p_2+p_3=0.
\label{SU3c}
\ee
Again the issue is whether some choice of the normalisation $c$
might simplify matters. With our choice of (\ref{cfun})
for example, the third term in (\ref{SU3cc}) turns into just
$v\lambda$, but no drastic simplification occurs in the constant
term unless at the special point $\epsilon = \frac{\pi}{2}$
(where $\wp'(\epsilon =\pi/2) = 0$) when -- modulo $c$ -- the whole
equation becomes linear in $x = \wp(\xi)$.

Simplifications do however take place in the perturbative limit
which for our purposes suffices to establish the prepotential.
In this limit we shall establish a convenient factorisation of our
curves. We note similar simplifications were found in
\cite{DP, vaninsky} in the  study of the trigonometric limit of
the elliptic Calogero-Moser models.
Indeed, in the  perturbative limit {\it and} at the special point
$\epsilon =\pi/2$, which was the context of \cite{BMMM1}, quite
dramatic simplifications occur. We shall study this case first, before
turning to the case of general $\epsilon$.  Before performing
this analysis for general $N$ however, let us consider the
simplifications for the example of (\ref{SU3cc}) given above.

In perturbative limit and at the special point
$\epsilon =\pi/2$ (\ref{SU3cc}) turns into
\be
\lambda^3 + v\lambda = (u\lambda^2 + 1)\tan \xi
\label{SU3tan}
\ee
for the choice $c=c_0$. Equivalently,
\be
\tan \xi = \frac{\lambda(\lambda^2 + v)}
{u\lambda^2 + 1}.
\label{SU3cot1}
\ee
This may be readily compared with the rational spectral curve
from \cite{BMMM1}. To make this comparison more immediate, let us
slightly change the choice of $c$: $c=-ic_0$. With this choice we get
\be
i\tan \xi = \frac{\lambda(\lambda^2 + v)}
{u\lambda^2 + 1}.
\label{SU3cot}
\ee
and should compare it with the curve from \cite{BMMM1}:
\be
w = -\frac{(\lambda -  \lambda_1)(\lambda - \lambda_2)(\lambda
-  \lambda_3)}
{( \lambda + \lambda_1)( \lambda + \lambda_2)( \lambda+ \lambda_3)}.
\label{SU3B}
\ee
Again, (\ref{SU3cot}) is equivalent to (\ref{SU3B}) after the
identification (\ref{wdef}), $w = -e^{2i\xi}$ and
with $u = \lambda_1 + \lambda_2 + \lambda_3$
and $v =  \lambda_1\lambda_2 + \lambda_2\lambda_3 + \lambda_3\lambda_1
= \lambda_1\lambda_2\lambda_3
\left(\lambda_1^{-1} + \lambda_2^{-1} + \lambda_3^{-1}\right)$,
$\lambda_1\lambda_2\lambda_3=1$.

\subsection{The special point $\epsilon=\omega =\pi/2$}

Here we  separately consider the special point where $\epsilon$
equals the real half-period $\omega=\pi/2$. We shall see that
simplifications occur even before taking the perturbative limit.
Using the fact that
$\sigma(\xi+2\omega)=-\sigma(\xi)e^{2\eta_1(\xi+\omega)}$
with $\eta_1 = \zeta(\omega)$ \cite{BE, WW}, we find that
\be
\frac{\sigma(\xi)\sigma(\omega)}{\sigma(\xi + \omega)}\equiv
c_+(\xi|\omega) = \alpha c_0(\xi|\omega) = \alpha^2 c_-(\xi|\omega)
\ee
with $\alpha = e^{-\xi\eta_1}$.
Then
\be
D_k(\xi|\epsilon)=\Bigg\{
\begin{array}{ll}
c_0\sp{-k} {\cal H}_1\sp{-k(k-2)/2}&k\quad{\rm even},\\
c_0\sp{-k} {\cal H}_1\sp{-k(k-2)/2} c_0 {\cal H}_1\sp{-1/2}&
k\quad{\rm odd}.
\end{array}
\ee
Here ${\cal H}_1=2 e_1\sp2 +e_2 e_3$ is independent of $\xi$
and we note that in the trigonometric limit ${\cal H}_1=1$.
The choice $c\sim c_0$ now essentially removes all $\xi$-dependence
in the spectral curve apart from a term linear
in $c_0$ that multiplies the odd  Hamiltonians. Similar to the $SU(3)$ case,
we choose $c=-ic_0$.
By absorbing
the (inessential) constant factors into the Hamiltonians,
$h_k= i^k{\cal H}_1\sp{-k(k-2)/2} H_k $
for even $k$ and
$h_k = i^{k-1}{\cal H}_1\sp{-1/2}{\cal H}_1\sp{-k(k-2)/2} H_k $
for odd $k$, one obtains the non-perturbative spectral curve
(\ref{Rsc}) at the special point $\epsilon = \omega$:
\be\label{spsc1}
ic_0^{-1}(\xi|\omega)=i\sqrt{\wp (\xi)-\wp (\omega)}=\frac
{\sum^N_{odd\ k}h_k(-\lambda)^{N-k}}
{\sum^N_{even\ k} h_k(-\lambda)^{N-k}}={P(\lambda)
-(-)^NP(-\lambda)\over P(\lambda)+(-)^NP(-\lambda)},
\ee
where
\be
P(\lambda)=\sum_{k}h_k(-\lambda)^{N-k}=\lambda^N+\ldots +1
\ee
These formulae may also be easily expressed in terms of the
Jacobi elliptic functions.  For example,
$\wp (\xi)-\wp(\omega)=(e_1-e_3){cn^2(u,k)\over sn^2(u,k)}$
and, therefore,
$c_0(\xi|\omega)=
{1\over \sqrt{e_1-e_3}}{sn(u,k)\over cn(u,k)}$; here $sn(u,k)$ and $cn(u,k)$
are the Jacobi functions \cite{BE, WW},
$u=(e_1-e_3)^{1/2} \xi$, $k$ is elliptic modulus.

We see then that substantial simplifications occur at the
special point $\epsilon = \omega$ before even considering the
perturbative limit. Now in the perturbative limit
$\sigma(z)\rightarrow \sin(z)e\sp{z\sp2 /6}$.
Using this we easily obtain that
\be
i\cot\xi={P(\lambda)-(-)^NP(-\lambda)\over P(\lambda)+
(-)^NP(-\lambda)},\qquad{\rm and}\qquad
w\equiv -e^{2i\xi}=(-)^{N}{P(\lambda)\over P(-\lambda)}.
\ee
With $dS\sim \log\lambda {dw\over w}$ this exactly reproduces
the curves of ref.\cite{BMMM1} where the prepotential (\ref{pp})
was calculated.

\subsection{The perturbative result for generic $\epsilon$}

We finally turn to the case of generic $\epsilon$ in the
perturbative limit.  The system in this limit is described by the
trigonometric \R model. We have already seen that in this limit
$\sigma$-functions are proportional to sines with
exponential factors. However, the periodicity of
our spectral curve means that these exponential factors
must cancel amongst themselves.
For example we find that in this limit
\be
D_k(\xi|\epsilon) =
 \frac{1}{\bar c_-\sp{k}}. \frac{\sin(\xi + (k-1)\epsilon)}{
\sin(\xi - \epsilon)}.\frac{(-1)\sp{k(k-1)\over 2}}
{\sin^{k(k-2)}(\epsilon)},
\label{Dfuntrig}
\ee
where $\bar c_-$ is given by (\ref{ctrig}).
Now the ratio
\be
\frac{\sin(\xi + (k-1)\epsilon)}{
\sin(\xi - \epsilon)}
=\cos k\epsilon + \sin k\epsilon \cot(\xi - \epsilon)
\label{fractocot}
\ee
is expressible in terms of the single function
$\cot(\xi - \epsilon)$.
This simple observation
enables us to  separate the variables $\xi$
and $\lambda$ in the equation for the spectral curve
upon choosing\footnote{
Note that this separation is not in general  possible
non-perturbatively where there is no analogue of eq.(\ref{fractocot}).
(This is because the ratios
$\frac{\sigma(\xi + (k-1)\epsilon)}{\sigma(\xi-\epsilon)}$ transform
differently under
$\xi-\epsilon \rightarrow \xi-\epsilon + 1$ and
$\xi-\epsilon \rightarrow \xi-\epsilon + \tau$ for different $k$.)
The two notable exceptions are:

(i) The case of $SU(2)$, when the variables  separated
for the standard choice $c = c_0$, see eq.(\ref{SU2c1});

(ii) The case of $\epsilon = \pi/2$ (for any $SU(N)$),
when separation of variables survives non-perturbatively,
see eq.(\ref{spsc1}).
}
  $c =-ie^{i\epsilon}\bar c_-$.
With $\bar h_k= {H_k}/ {\sin^{k(k-2)}(\epsilon)}$
we may simplify the spectral curve to give
\be
i\cot(\xi - \epsilon)=
-\frac{\sum_{odd\ k}^N {\bar h_k}
(-\lambda)^{N-k}\left(e^{2ik\epsilon}+1\right)}
{\sum_{even\ k}^N {\bar h_k}
(-\lambda)^{N-k}\left(e^{2ik\epsilon}-1\right)}=
\frac{P(\lambda)+e^{2iN\epsilon}P(\lambda
e^{-2i\epsilon})}{P(\lambda)-e^{2i\epsilon N}P(\lambda
e^{-2i\epsilon})}
\ee
where $P(\lambda)=\sum_{all\ k}^N \bar h_k(-\lambda)^{N-k}=
\prod_i^N(\lambda -e^{2ia_i})$ with some constants $a_i$,
$\sum_i^N a_i=0$. Introducing the variable $w=e^{2i(\xi-\epsilon)}$,
one finally arrives the spectral curve in the form
\be\label{scf}
w=e^{-2i\epsilon N}\frac{P(\lambda)}{P(\lambda e^{-2i\epsilon})}
\ee
with
\be\label{dSc}
dS\cong\log\lambda {dw\over w}.
\ee

Thus, we have shown that our system leads in the perturbative limit
to the rational spectral curve
(\ref{scf}) and the generating differential (\ref{dSc}).
One may now calculate the corresponding prepotential using
the residue formula. This calculation has been done quite generally
in \cite{MM}, where the residue formula is applied to a general
rational function of the form
$$
w\sim \frac{\prod_i\sp{N_c}(\lambda-\lambda_i)}{\sqrt{\prod_\alpha
\sp{N_f}(\lambda-\lambda_\alpha)}},
$$
with $\lambda_i=e\sp{2 a_i}$, $\sum_i a_i =0$, $\lambda_\alpha=
e\sp{2 m_\alpha}$. By choosing $N_c=N$ and $N_f=2 N$, with the
hypermultiplets  masses pairwise coinciding and
equal to $a_i+\epsilon$ we obtain our curve (\ref{scf}).
One then finds the prepotential  from (3.37) of \cite{MM}
gives the stated prepotential (\ref{pp}), and we are done.
Further, upon setting $\epsilon=\pi/2$ we reproduce the results
of the previous subsection.

\section{Conclusion}

In the context of \SW we have analysed some properties of the
most general compactification from five dimensions to four
dimensions with all of the fields belonging to adjoint representation
of the gauge group $SU(N)$.
Here \SW describes the relevant low-energy effective action
in terms of the finite-dimensional elliptic \R integrable model,
of which the elliptic Calogero, relativistic and ordinary Toda
chains are particular limits.
Our work unifies these previous treatments.
Special attention was devoted to the perturbative (weak coupling)
limit, when $\tau \rightarrow i\infty$ and the elliptic models
degenerate into trigonometric ones.
Two topics are still beyond this general model: compactification
from six dimensions and the inclusion of supermultiplets in the
fundamental representation of the gauge group.
For both these cases the elliptic Sklyanin (XYZ) spin-chain model
\cite{XYZ} seems to be relevant (see arguments in \cite{ggm2,MM}
and \cite{fund} respectively).
In the present paper the \R model has been discussed
with the help of an $N\times N$ Lax operator and we did not address the
issue of a possible spin-chain-like description.
This is an area warranting further investigation.
The most immediate difficulty in extending this work to other groups
is the lack of a suitable Lax formulation for general root
systems, and we also highlight this as an interesting problem.

Another open question concerns the generalized WDVV equations
\cite{WDVV}.
At the special point $\epsilon = \pi/2$ these are known to hold
\cite{BMMM1}, at least in the perturbative limit.
Equally, they are known to {\it fail} \cite{rev, WDVV2}
for the  perturbative limit of the elliptic Calogero model,
that is the trigonometric Calogero-Sutherland system.
This implies that there can be problems with the naive WDVV equations
for the \R system under consideration, unless the additional moduli
$\tau$ and $\epsilon$ are cleverly taken into account.
It may indeed be more illuminating to consider the more general \R
setting rather than that of the Calogero system, especially given
the result of \cite{BMMM1} at $\epsilon = \pi/2$.
Perhaps the most intriguing property of these models
is that perturbatively they are always described by the
rational curves we have exhibited in this paper. Nevertheless, even
at the
perturbative level, the WDVV equations have only been established
at the special
point $\epsilon=\pi/2$.  This issue deserves further analysis.

\section{Acknowledgements}

We acknowledge discussions with A.Gorsky, S.Kharchev, N.Nekrasov and
A.Zabrodin.

A.Mar., A.Mir. and A.Mor. are grateful for the hospitality
of University of Edinburgh and the support of EPSRC (grant GR/M08134).
A.Mir. and H.W.B. also acknowledge the Royal Society for
support under a joint project.

Our research is partly supported by the
RFBR grants 98-01-00344 (A.Mar.), 98-01-00328 (A.Mir.),
98-02-16575 (A.Mor.), INTAS grants 96-482 (A.Mar.), 97-0103 (A.Mir.),
the program for support of the scientific schools 96-15-96798
(A.Mir.) and the Russian President's grant 96-15-96939 (A.Mor.).

\end{document}